\documentclass[english]{paper}
\usepackage[latin9]{inputenc}
\usepackage{geometry}
\usepackage{rotfloat}
\usepackage{amsthm}
\usepackage{amsmath}
\usepackage{amssymb}
\usepackage{graphicx}
\usepackage[authoryear]{natbib}

\makeatletter

\providecommand{\tabularnewline}{\\}

\numberwithin{equation}{section}
\numberwithin{figure}{section}
  \theoremstyle{definition}
  \newtheorem*{example*}{\protect\examplename}
  \theoremstyle{remark}
  \newtheorem*{claim*}{\protect\claimname}

\makeatother

\usepackage{babel}
  \providecommand{\claimname}{Claim}
  \providecommand{\examplename}{Example}

\begin{document}

\title{Keeping greed good: sparse regression under design uncertainty with
application to biomass characterization}

\author{David J. Biagioni, %
\thanks{Corresponding author: biagioni@colorado.edu. Department of Applied
Mathematics, UCB 526, University of Colorado at Boulder , Boulder,
CO 80309-0526.%
} %
\thanks{Computational Sciences Center, National Renewable Energy Laboratory,
16253 Denver West Parkway, Golden, CO.%
} \and Ryan Elmore, \footnotemark[2] \and Wesley Jones \footnotemark[2]}
\maketitle
\begin{abstract}
\ \ In this paper, we consider the classic measurement error regression
scenario in which our independent, or design, variables are observed
with several sources of additive noise. We will show that our motivating
example's replicated measurements on both the design and dependent
variables may be leveraged to enhance a sparse regression algorithm.
Specifically, we estimate the variance and use it to scale our design
variables. We demonstrate the efficacy of scaling from several points
of view and validate it empirically with a biomass characterization
data set using two of the most widely used sparse algorithms: least
angle regression (LARS) and the Dantzig selector (DS). 
\end{abstract}

\section{Introduction\label{sec:Introduction}}

This paper is motivated by the practical problem of how to meaningfully
perform sparse regression when the predictor variables are observed
with measurement error or some source of uncertainty. We will refer
to this error or noise as design uncertainty to emphasize that perturbations
in the design matrix may arise from a number of random sources unrelated
to experimental or measurement error \emph{per se}. Recent work in
this area has just begun to address the issue of sparse regression
under design uncertainty from a theoretical point of view. We are
primarily interested in describing an approach that, while theoretically
justifiable, is essentially pragmatic and broadly applicable. In short,
we argue that greed - a basic feature of many sparsity promoting algorithms
- is indeed good \citep{TROPP-2004}, so long as the design data is
scaled by the uncertainty variances. We demonstrate the efficacy of
scaling from several points of view and validate it empirically with
a biomass characterization data set using two of the most widely used
sparse algorithms: least angle regression (LARS) and the Dantzig selector
(DS).

Our work was motivated by an example from a biomass characterization
experiment related to work at the National Renewable Energy Laboratory.
The example is described in detail in Section \ref{sec:Application-to-biomass}
and contains repeated measurements of mass spectral (design, or predictor)
and sugar mass fraction (response) values within each switchgrass
sample. The domain scientists' goal was to find a small subset of
masses in the spectrum that could be used to predict sugar mass fraction.
We will show that the replication of each measurement allows for simple
estimates of the error variances which, in turn, may be used to guide
the model selection procedure. Thus, we are interested in sparse regression
under design uncertainty. We would also like for a scientist examining
the model to have some hope of interpreting its meaning, either for
immediate understanding or to indicate new research directions.

Sparse regression by $l_{1}$ minimization is a thriving and relatively
young field of research. In the statistical inference literature,
early stepwise-type algorithms paved the way for the now-familiar
lasso \citep{TIBSHI-1996.pdf}, least angle regression (LARS) \citep{E-H-J-T-2004},
and many variants tailored to specific problems (for example, \citet{YUA-LIN-2007,ZOU-HAS-2005,TIB-TAY-2011,H-T-T-W-2007.pdf,P-R-R-W-2011}).
A parallel evolution in the signal processing literature led to the
development of widely used basis and matching pursuit algorithms \citep{CH-DO-SA-1998,CH-DO-SA-2001,TROPP-2004},
the Dantzig selector (DS) \citep{CAN-TAO-2007}, and many others (see, \emph{e.g.},
\citet{ELAD-2010}, Chapters 3 and 5, for a good overview). Despite
their mostly independent development, the algorithms coming out of
the statistical and signal processing worlds lead to remarkably similar
results in many applications (\emph{e.g.}, \citet{BI-RI-TS-2009}).

Linear regression under the assumption of design uncertainty has,
in comparison, a long history, going by various names such as error
in variables or functional modeling, and a variety of techniques have
been developed to address it (\emph{e.g.}, \citet{GILLAR-2010,FULLER-1987,FULLER-1995}).
Until fairly recently, however, much of the analysis of sparse representations
has not confronted this issue. As we will discuss, there is good reason
for this, namely, that this problem obfuscates the goal of sparse regression.

Several recent works that have looked at sparse regression under various
assumptions about the noise should be mentioned. \citet{ROS-TSY-2010},
develop a Dantzig-like estimator that they argue is more stable than
the standard lasso or Dantzig. \citet{SUN-ZHA-2011} describe an algorithm
to estimate the lasso solution and the noise level simultaneously.
A similar idea, leading to the ``adaptive lasso'', was developed
by \citet{HU-MA-ZH-2008} under homoscedastic assumptions. An algorithm
that hybridizes total least squares \citep{GOL-VAN-1980}, a computational
error in variables model, and the lasso was also recently published
by \citet{ZH-LE-GI-2011}.

The work that comes nearest to our discussion is by \citet{WAG-DET-2011,WAG-DET-2011b}.
In these papers, the authors present some asymptotic results for bridge
and lasso estimators under the assumption of heteroscedasticity. In
particular, they develop a weighting scheme that leads to adaptive
lasso estimates that are sign consistent (\emph{i.e.}, they satisfy
the ``oracle property'').

We consider this paper to be somewhat disjoint from the aforementioned
for two reasons. First, we are primarily concerned with an approach
that incorporates empirical knowledge of design uncertainty into the
analysis. Second, we wish to argue from a more general, and necessarily
more heuristic, point of view that does not require stringent conditions,
such as those described by \citet{WAG-DET-2011}, Section 3, to hold.
In other words, we want to allow for the possibility that the data
that is given to us may be ``messy.'' For example, we do not expect
the design matrix to satisfy the restricted isometry property or to
have low mutual coherence which, under certain circumstances, would
guarantee the efficacy of an appropriate sparse algorithm.

A central notion throughout this paper is that many of the standard
sparse regression algorithms are greedy, that is, they search for
a solution incrementally, using the best available update at any given
point in the search. As such, we argue that estimates of uncertainty
should modify the notion of greed. Some algorithms, such as orthogonal
matching pursuit (OMP), basis pursuit (BP), and forward stagewise
regression (FS), are explicitly greedy. Others, like those that solve
the lasso and Dantzig selector problems, may also be viewed as greedy
via their connection to homotopy methods \citep{ASI-ROM-2009,ASI-ROM-2010,E-H-J-T-2004}.
These methods generally take an initial estimate of the solution and
move along a continuous path to the final one, choosing the best available
search direction at each step.

Initially, we take forward stagewise (FS) regression as our prototype
for analysis, noting its close relationship to the lasso and LARS
\citep{H-T-T-W-2007.pdf}, as well as OMP and BP \citep{ELAD-2010}.
We show that for all solution paths of a fixed norm, the uncertainty
of the residual and the solution norm have a dual-like relationship
in which the homogeneity of one induces inhomogeneity of the other,
and that one can move from one problem to the other via a scaling
of the design variables. From the standpoint of sparse pursuit, we
argue that, as a general principle, uniform growth of the uncertainty
along the solution path is preferable to uniform growth of the solution
norm. Similar arguments are shown to apply to the Dantzig selector
(DS). We then compare LARS and DS cross-validated model selection
on a repeated measures biomass characterization data set in which
variances are estimated via an analysis of variance (ANOVA) model.
In this application, scaling by the uncertainty variance leads to
sparser and more accurate models. Prediction error is reduced even
further if, after down-selection of the variables by LARS and DS,
the solution is updated via an $l_{2}$ method such as ridge regression.

\section{Regression under design uncertainty}

In this section, we formulate the model of interest and outline the
challenges posed by design uncertainty. More importantly, we derive
a simple estimate of this quantity which will play a central role
in the discussion. We also give a simple example that illustrates
how a very sparse solution can sometimes be associated with more of
the design uncertainty than a less sparse one, further motivating
our approach.

\subsection{Model\label{sub:Model}}

We consider response data of the form, 
\begin{eqnarray}
y_{i} & = & w_{i}+\epsilon_{i},\label{eq:y-model}\\
w_{i} & \sim & N(0,\sigma_{w}^{2}),\nonumber \\
\epsilon_{i} & \sim & N(0,\sigma_{\epsilon}^{2}),\nonumber \\
\mbox{Cov}(\boldsymbol{w},\boldsymbol{\epsilon}) & = & 0,\nonumber 
\end{eqnarray}
and design data, 
\begin{eqnarray}
x_{ij} & = & v_{ij}+\delta_{ij},\label{eq:x-model}\\
v_{ij} & \sim & N(0,\sigma_{v_{j}}^{2}),\nonumber \\
\delta_{ij} & \sim & N(0,\sigma_{\delta_{j}}^{2}),\nonumber \\
\mbox{Cov}(\boldsymbol{v}_{j},\boldsymbol{\delta}_{j}) & = & 0,\,\,\,\,\,\,\,\,\,\,\,\,\forall j\nonumber 
\end{eqnarray}
for $1\leq i\leq n$ and $1\leq j\leq p$. The assumptions on $\mathbf{w}$
and $\mathbf{v}_{j}$ imply that the data are mean centered, and we
interpret $\boldsymbol{\epsilon}$ and $\boldsymbol{\delta}_{j}$
as independent uncertainties, 
\begin{eqnarray*}
\mbox{Cov}(\boldsymbol{\epsilon},\boldsymbol{\delta}_{j}) & = & 0,\,\,\,\forall j,\\
\mbox{Cov}(\boldsymbol{\delta}_{j},\boldsymbol{\delta}_{k}) & = & 0,\,\,\, j\neq k,
\end{eqnarray*}
arising from measurement error, natural within-sample variability,
or other random sources.

We will often express the system in matrix form, 
\begin{eqnarray*}
\mathbf{y} & = & \mathbf{w}+\boldsymbol{\epsilon},\\
X & = & V+\Delta,
\end{eqnarray*}
where $\boldsymbol{\epsilon}=(\epsilon_{1},\epsilon_{2},\dots,\epsilon_{n})'$
and $\Delta=[\delta_{ij}]_{n\times p}$, and take the columns of $X$
and $\mathbf{y}$ to be scaled to unit variance, leading to the constraints
\begin{eqnarray*}
\sigma_{w}^{2}+\sigma_{\epsilon}^{2} & = & 1,\\
\sigma_{v_{j}}^{2}+\sigma_{\delta_{j}}^{2} & = & 1,\,\,\,\,\,\forall j.
\end{eqnarray*}
Furthermore, we have 
\begin{eqnarray*}
\mbox{Cov}(X) & = & \mbox{Cov}(V)+\Sigma^{2},
\end{eqnarray*}
where $\Sigma^{2}\equiv\mbox{Cov}(\Delta)=\mbox{diag}(\sigma_{\delta_{1}}^{2},\sigma_{\delta_{2}}^{2},\dots,\sigma_{\delta_{p}}^{2})$
by the independence of the errors. When using finite sample estimates
$s_{\delta_{j}}^{2}$ of $\sigma_{\delta_{j}}^{2}$, we denote the
corresponding matrix by $S^{2}$. In the absence of noise ($\sigma_{\epsilon}^{2}=0$
and $\sigma_{\delta_{j}}^{2}=0,\,\forall j$), we assume that the
design and response admit a linear model, 
\begin{equation}
\mathbf{w}=V\boldsymbol{\beta}.\label{eq:linear-model}
\end{equation}
We are particularly interested in the case where $\boldsymbol{\beta}$
is sparse: loosely speaking, many of its elements are zero.

In the application discussed in Section \ref{sec:Application-to-biomass},
repeated measurements are used to estimate the variances $\sigma_{\delta_{j}}^{2}$.
For a more theoretical application, it may be the case that these
parameters are known exactly. Either way, for the remainder of the
discussion we assume that either the variances or their sample estimates
are available.

\subsection{The challenge of design uncertainty\label{sub:The-challenge-of-heteroscedastic}}

One intrinsic challenge in working with noisy design data is that
the estimated regression coefficients are attenuated from their true
values. Suppose, instead of (\ref{eq:linear-model}), we were to solve
\begin{equation}
\mathbf{y}=X\boldsymbol{\beta}\label{eq:linear-model-with-noise}
\end{equation}
via ordinary least squares (OLS) to obtain $\hat{{\beta}}$. For $p=1$,
it is straightforward to show that 
\begin{equation}
\frac{\mathbb{E}\{\hat{\beta}\}}{\beta}=\frac{\sigma_{\nu}^{2}}{\sigma_{\nu}^{2}+\sigma_{\delta}^{2}}<1,\label{eq:reliability-ratio}
\end{equation}
where $\mathbb{E}$ denotes expected value. This implies that the
estimators are biased towards zero by an amount that depends on the
signal-to-noise ratio, $\sigma_{\nu}^{2}/\sigma_{\delta}^{2}$. More
generally, for any full rank $X\in\mathbb{R}^{n\times p}$ with $n\geq p$,
$X$ may be diagonalized such that in the new system of coordinates
an analogous result holds.

Design uncertainty also degrades the model fit even if the exact solution
$\boldsymbol{\beta}$ is known. To see this, consider the residual
error when design uncertainty is present: 
\begin{eqnarray}
\mathbb{E}\{||\mathbf{y}-X\boldsymbol{\beta}||_{2}^{2}\} & = & \mathbb{E}\{||(\mathbf{w}+\boldsymbol{\epsilon})-(V+\Delta)\boldsymbol{\beta}||_{2}^{2}\}\nonumber \\
 & = & \mathbb{E}\{||\boldsymbol{\epsilon}-\Delta\boldsymbol{\beta}||_{2}^{2}\}\nonumber \\
 & = & \mathbb{E}\{\boldsymbol{\epsilon}'\boldsymbol{\epsilon}-2\boldsymbol{\epsilon}'\Delta\boldsymbol{\beta}+\boldsymbol{\beta}'\Delta'\Delta\boldsymbol{\beta}\}\label{eq:residual-error}\\
 & = & \sigma_{\epsilon}^{2}+\boldsymbol{\beta}'\mathbb{E}\{\Delta'\Delta\}\boldsymbol{\beta}\nonumber \\
 & = & \sigma_{\epsilon}^{2}+\boldsymbol{\beta}'\Sigma^{2}\boldsymbol{\beta}\nonumber \\
 & = & \sigma_{\epsilon}^{2}+||\Sigma\boldsymbol{\beta}||_{2}^{2},\nonumber 
\end{eqnarray}
where we have explicitly used (\ref{eq:linear-model}) and the independence
of the errors.

This brings us to a main point, that the contribution of the design
uncertainty to the residual is of the form $||\Sigma\boldsymbol{\beta}||_{2}^{2}$,
which is quadratic in $\Sigma$. While we may only have access to
the attenuated estimate $\hat{\boldsymbol{\beta}}$ of $\boldsymbol{\beta}$,
the structure of the residual error remains the same with respect
to the error variances. We illustrate the effect this can have on
sparse regression with a simple example. 
\begin{example*}
Suppose $p=3$ and 
\begin{eqnarray*}
\mathbf{w} & = & v_{1},\\
v_{1} & = & \frac{1}{\sqrt{2}}(v_{2}+v_{3}),\\
\sigma_{\epsilon}^{2} & = & 0,\\
\Sigma^{2} & = & \mbox{diag}(\frac{1}{2},\frac{1}{4},\frac{1}{4}).
\end{eqnarray*}
The system admits the two solutions $\boldsymbol{\beta}^{(1)}=(1,0,0)'$
and $\boldsymbol{\beta}^{(2)}=\frac{1}{\sqrt{2}}(0,1,1)'$. The first
solution is the sparsest but in light of (\ref{eq:residual-error})
has greater expected error since $||\Sigma\boldsymbol{\beta}^{(1)}||_{2}=1/2$
while $||\Sigma\boldsymbol{\beta}^{(2)}||_{2}=\sqrt{2}/4$. Hence,
recovery of the sparsest solution results in greater uncertainty in
the fit than the less sparse one. The issue becomes even more prominent
- and more difficult to track - in higher dimensions with non-trivial
covariance of the design matrix. 
\end{example*}
Apparently,\emph{ greed is not always good under} \emph{design uncertainty}.

\section{Scaling penalizes design uncertainty in the solution path}

In this section, we briefly describe a prototypical greedy algorithm
for sparse regression, forward stagewise regression (FS). We do so
because it is helpful to have a particular algorithm in mind for the
discussion, and this one is particularly easy to understand. In addition,
it solves the widely-used lasso optimization problem and thus is closely
related to a variety of other important algorithms \citep{E-H-J-T-2004,H-T-T-W-2007.pdf}.
Next, we state the main result and provide simple algebraic and geometric
interpretations of it. Finally, we note implications of the result
for the Dantzig selector problem.

\subsection{A prototypical pursuit algorithm: forward stagewise (FS) regression}

The FS algorithm may be summarized as follows: 
\begin{enumerate}
\item Fix small $\gamma>0$ and initialize: $\hat{\boldsymbol{\beta}}=\mathbf{0}$,
$\mathbf{r}=\mathbf{y}$. 
\item Identify the design variable $\mathbf{x}_{j}$ most correlated with
$\mathbf{r}$. 
\item Incremental update%
\footnote{In LARS, the step is computed in a particularly efficient way but
the final solution path is essentially the same.%
} : $\hat{\beta}_{j}\leftarrow\hat{\beta}_{j}+\eta_{j}$, where $\eta_{j}=\gamma\cdot\mbox{sign}(\mbox{Corr}(\mathbf{x}_{j},\mathbf{r})).$ 
\item Subtract the projection of $\mathbf{r}$ onto $\mathbf{x}_{j}$: $\mathbf{r}\leftarrow\mathbf{r}-\eta_{j}\mathbf{x}_{j}$. 
\item If the residual norm is small enough, stop. Otherwise, return to step
2. 
\end{enumerate}
Qualitatively, the algorithm finds the best search direction - the
coordinate with highest residual correlation - and takes a small step
in that direction. It does so iteratively, updating the solution and
residual at each step, until the minimal residual error is reached.

As an optimization procedure, FS regression (like the lasso and LARS)
implicitly solves 
\begin{equation}
\arg\min_{\boldsymbol{\beta}\in\mathbb{R}^{p}}\frac{1}{2}||\mathbf{y}-X\boldsymbol{\beta}||_{2}^{2}\,\,\,\,\,\mbox{subject to}\,\,\,\,\,||\boldsymbol{\beta}||_{1}<\lambda,\label{eq:lasso-obj-fun-constrained-form}
\end{equation}
which is often expressed in Lagrangian form, 
\begin{equation}
\arg\min_{\boldsymbol{\beta}\in\mathbb{R}^{p}}\frac{1}{2}||\mathbf{y}-X\boldsymbol{\beta}||_{2}^{2}+\lambda||\boldsymbol{\beta}||_{1},\label{eq:lasso-obj-fun-lagrange-form}
\end{equation}
for a range of values of the tuning parameter $\lambda>0$. In the
limit $\lambda\rightarrow0$, the optimum is attained by the ordinary
least squares solution, while solutions for $\lambda\rightarrow\infty$
are increasingly sparse (\textbf{$\lambda=\infty\Leftrightarrow\boldsymbol{\beta}=0$})
\citep{TIBSHI-1996.pdf}.

\subsection{Main result}

Our main result is simple: it says that for all solutions of a fixed
norm, the accumulated design uncertainty (estimated by $||\Sigma\hat{\boldsymbol{\beta}}||_{2}$
in equation (\ref{eq:residual-error})) is path-dependent unless the
data are scaled by the uncertainty variance. In other words, scaling
the data leads to a uniform increase of the design uncertainty contribution,
independent of the search direction.

To see this (and with a slight abuse of notation), we first modify
equation (\ref{eq:linear-model-with-noise}) to include scaling of
the design variables, 
\begin{equation}
\mathbf{y}=XD^{-1}\boldsymbol{\beta},\label{eq:linear-model-with-scaling}
\end{equation}
noting that if $\boldsymbol{\beta}$ solves (\ref{eq:linear-model}),
then $D\boldsymbol{\beta}$ solves (\ref{eq:linear-model-with-scaling}).
The expected residual variance is then 
\begin{eqnarray}
\mathbb{E}\{||\mathbf{y}-XD^{-1}\boldsymbol{\beta}||_{2}^{2}\} & = & \sigma_{\epsilon}^{2}+\mathbb{E}\{\boldsymbol{\beta}'D^{-1}\Delta'\Delta D^{-1}\boldsymbol{\beta}\}\nonumber \\
 & = & \sigma_{\epsilon}^{2}+\boldsymbol{\beta}'D^{-1}\Sigma^{2}D^{-1}\boldsymbol{\beta}\nonumber \\
 & = & \sigma_{\epsilon}^{2}+||\Sigma D^{-1}\boldsymbol{\beta}||_{2}^{2}.\label{eq:residual-error-with-scaling}
\end{eqnarray}

Now let $U(\boldsymbol{\beta};D)=||\Sigma D^{-1}\boldsymbol{\beta}||_{2}^{2}$
denote the design uncertainty denoted with a solution $\boldsymbol{\beta}$
of fixed norm, $||\boldsymbol{\beta}||_{2}^{2}=T^{2}$. Clearly, the
uncertainty is independent of the uncertainty variances when $D=\Sigma^{-1}$
(\emph{i.e.,} $D_{jj}=\sigma_{\delta_{j}}^{-1}$). Specifically, $U(\boldsymbol{\beta};D)=T^{2}$
for any $\Sigma$.

Scaling the data will result in solutions of different norms, so that
two solutions of norm $T$ under different scalings $D_{1}^{-1}$
and $D_{2}^{-1}$ are not directly comparable in terms of the underlying
optimization problem. However, the result says that scaling by $D=\Sigma$
leads to a solution space in which all solutions of identical norm
have identical uncertainty.

\subsubsection{Algebraic interpretation of scaling\label{sub:Algebraic-interpretation} }

Based on our claim, we consider scaling the columns of $X$ by the
associated uncertainties, 
\begin{eqnarray*}
D & = & \Sigma,\\
X & \leftarrow & XD^{-1}.
\end{eqnarray*}
The most obvious effect of scaling is that the correlations change
and so (potentially) does the order in which the variables are selected
(step 2 of the FS algorithm). Recalling that the columns of $X$ and
$\mathbf{y}$ have unit variance, we initially have 
\[
\mbox{Corr}(\mathbf{x}_{j},\mathbf{y})=\mathbf{x}_{j}'\mathbf{y},
\]
while after scaling, 
\[
\mbox{Corr}(\mathbf{x}_{j},\mathbf{y})=\mathbf{x}_{j}'\mathbf{y}/\sigma_{\delta_{j}}.
\]

A less obvious effect of scaling is that the underlying problem (\ref{eq:lasso-obj-fun-lagrange-form})
is transformed so that uncertainty in the solution path is penalized
explicitly. The scaled problem, 
\begin{equation}
\arg\min_{\boldsymbol{\beta}\in\mathbb{R}^{p}}\frac{1}{2}||\mathbf{y}-XD^{-1}\boldsymbol{\beta}||_{2}^{2}+\lambda||\boldsymbol{\beta}||_{1},\label{eq:scaled-lasso}
\end{equation}
by a simple change of variables, $\boldsymbol{\beta}\leftarrow D\boldsymbol{\beta}$,
may be written

\[
\arg\min_{\boldsymbol{\beta}\in\mathbb{R}^{p}}\frac{1}{2}||\mathbf{y}-X\boldsymbol{\beta}||_{2}^{2}+\lambda||D\boldsymbol{\beta}||_{1}.
\]
We note that this is the ``generalized lasso'' problem described
in \citet{TIB-TAY-2011}. The lasso penalty term represents the ``$l_{1}$
version'' of the design uncertainty (recall that $||\Sigma\boldsymbol{\beta}||_{2}\leq||\Sigma\boldsymbol{\beta}||_{1}\leq\sqrt{p}||\Sigma\boldsymbol{\beta}||_{2},$
by norm equivalence). Hence, scaling by $D=\Sigma$ leads to a direct
$l_{1}$ penalization of design uncertainty within the lasso framework.

\subsubsection{Geometric interpretation of scaling\label{sub:Geometric-interpretation}}

\begin{figure}
\begin{centering}
\includegraphics[scale=0.5]{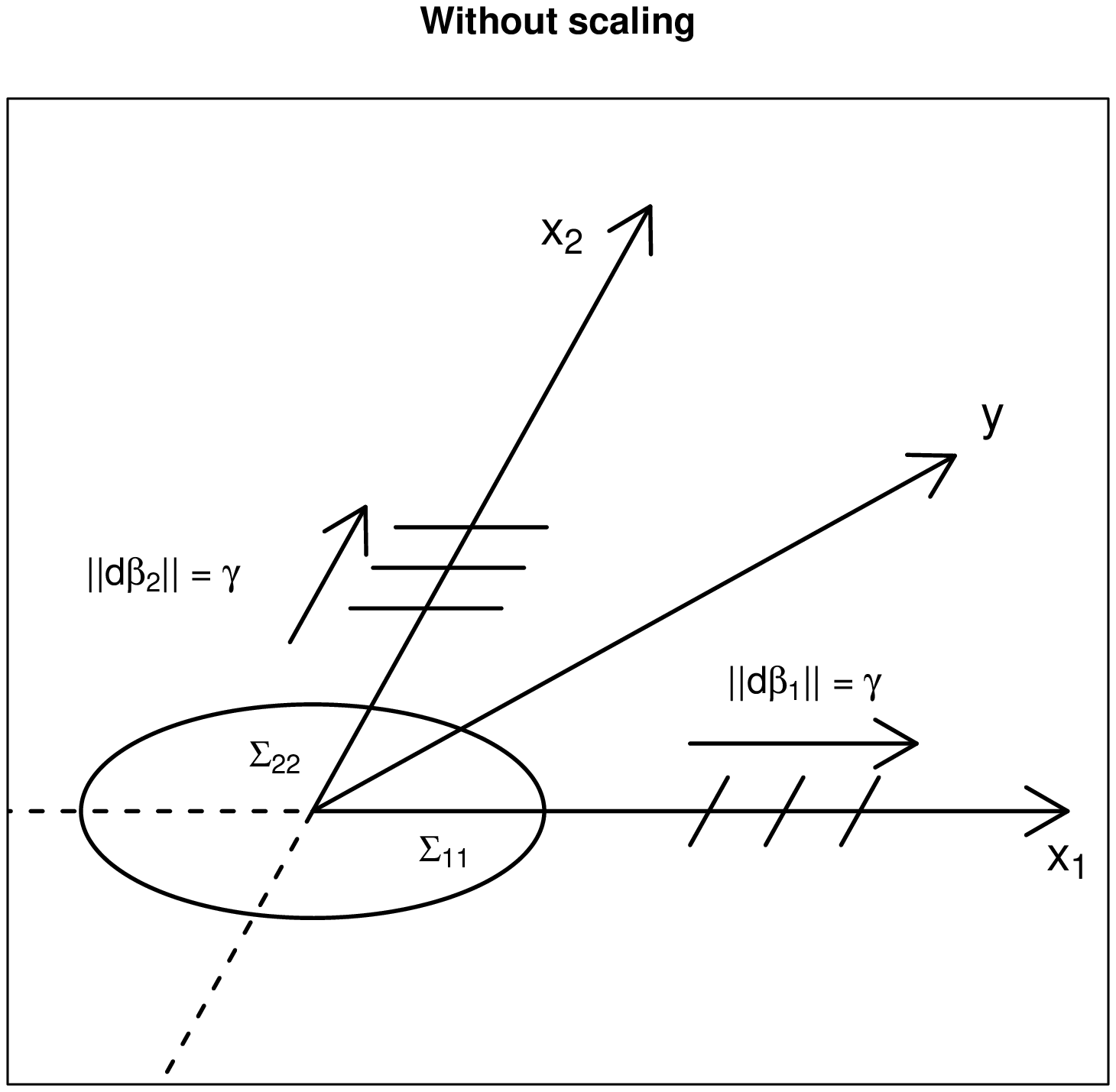} \\
 \includegraphics[scale=0.5]{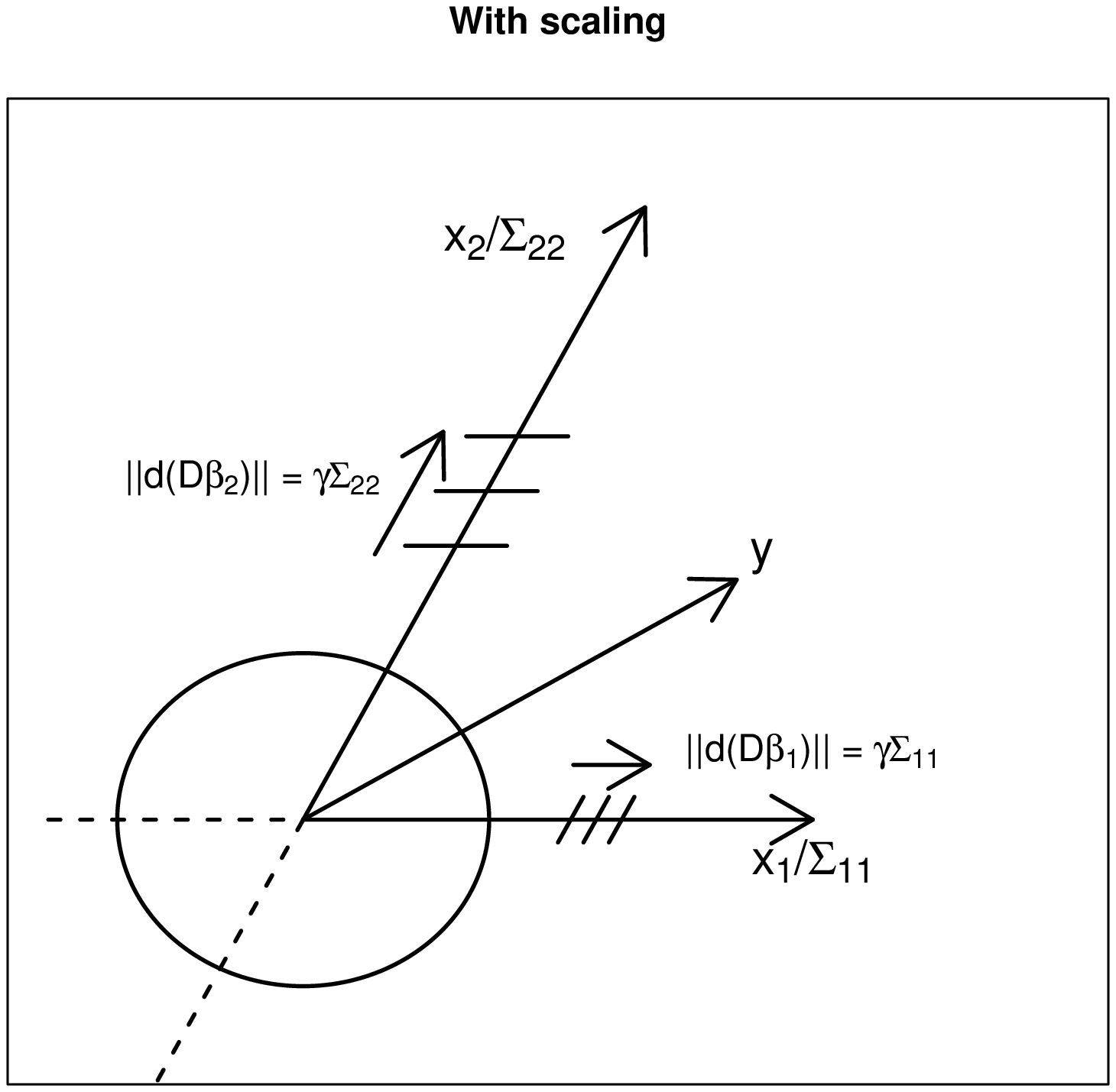} 
\par\end{centering}

\caption{The dual-like relationship between uncertainty and solution norm.
Left: Without scaling, uncertainty grows anisotropically (indicated
by an ellipse around the origin) for fixed step size $\gamma$ of
the greedy algorithm, while the solution norm grows uniformly (indicated
by perpendicular contour lines on each axis). Right: After scaling
by $D^{-1}$, the uniformity of uncertainty and solution norm are
reversed. The notation $||d(\cdot)||_{1}$ represents the increase
in the $l_{1}$ penalty term for a small step in the solution path.
\label{fig:geometry-of-scaling}}
\end{figure}

Geometrically, scaling by the uncertainty induces a dual-like problem
in which the homogeneity of solution norm and the uncertainty are
reversed (Figure \ref{fig:geometry-of-scaling}). In particular, before
scaling, a step of fixed size $\gamma$ leads to constant growth of
the $l_{1}$ penalty but potentially non-uniform growth of the uncertainty.
After scaling, on the other hand, the uncertainty grows uniformly
at each step while the $l_{1}$ penalty does not.

Of course, for a given data set, the greedy algorithms we have discussed
are not random but deterministic. But if we consider the task of sparse
regression as applying to an ensemble of noisy data sets, one can
think of the solution paths as being effectively random (for a similar
line of reasoning see, \emph{e.g.}, \citet{DON-TSA-2008}). That is,
a statistical analysis of the algorithm is then necessarily and justifiably
carried out in terms of expectations, rather than specific search
paths.

\subsection{Connection to the Dantzig selector}

While a detailed analysis is beyond our scope, we take a brief moment
to point out the connection between scaling and the Dantzig selector.
\citet{CAN-TAO-2007} proposed an alternative formulation for sparse
regression, 
\[
\arg\min_{\boldsymbol{\beta}}||\boldsymbol{\beta}||_{1}\,\,\,\,\,\mbox{subject to}\,\,\,\,\,||X'(\mathbf{y}-X\boldsymbol{\beta})||_{\infty}\leq\lambda\sigma_{\epsilon},
\]
where $\lambda>0$ is a tuning parameter (different from the lasso
parameter) and $\sigma_{\epsilon}^{2}$ is the variance in (\ref{eq:y-model})
. The Dantzig selector has two main features that distinguish it from
other pursuit algorithms. The first is that the problem may be written
explicitly as a linear program (LP), for instance, 
\begin{eqnarray*}
\arg\min_{\boldsymbol{\alpha},\boldsymbol{\beta}}\boldsymbol{1}'\boldsymbol{\alpha} & \,\,\mbox{subject to}\,\, & -\boldsymbol{\alpha}\leq\boldsymbol{\beta}\leq\boldsymbol{\alpha}\\
 & \mbox{and} & -\sigma\lambda\mathbf{1}\leq X'(\mathbf{y}-X\boldsymbol{\beta})\leq\sigma\lambda\mathbf{1}.
\end{eqnarray*}
The second is that the $l_{\infty}$ constraint is with respect to
residual \emph{correlations} as opposed to residual \emph{error}.
This seems intuitively correct since, in the presence of noise, we
would expect the residual corresponding to an optimal solution to
have exactly this property.

\begin{figure}
\begin{centering}
\includegraphics[scale=0.5]{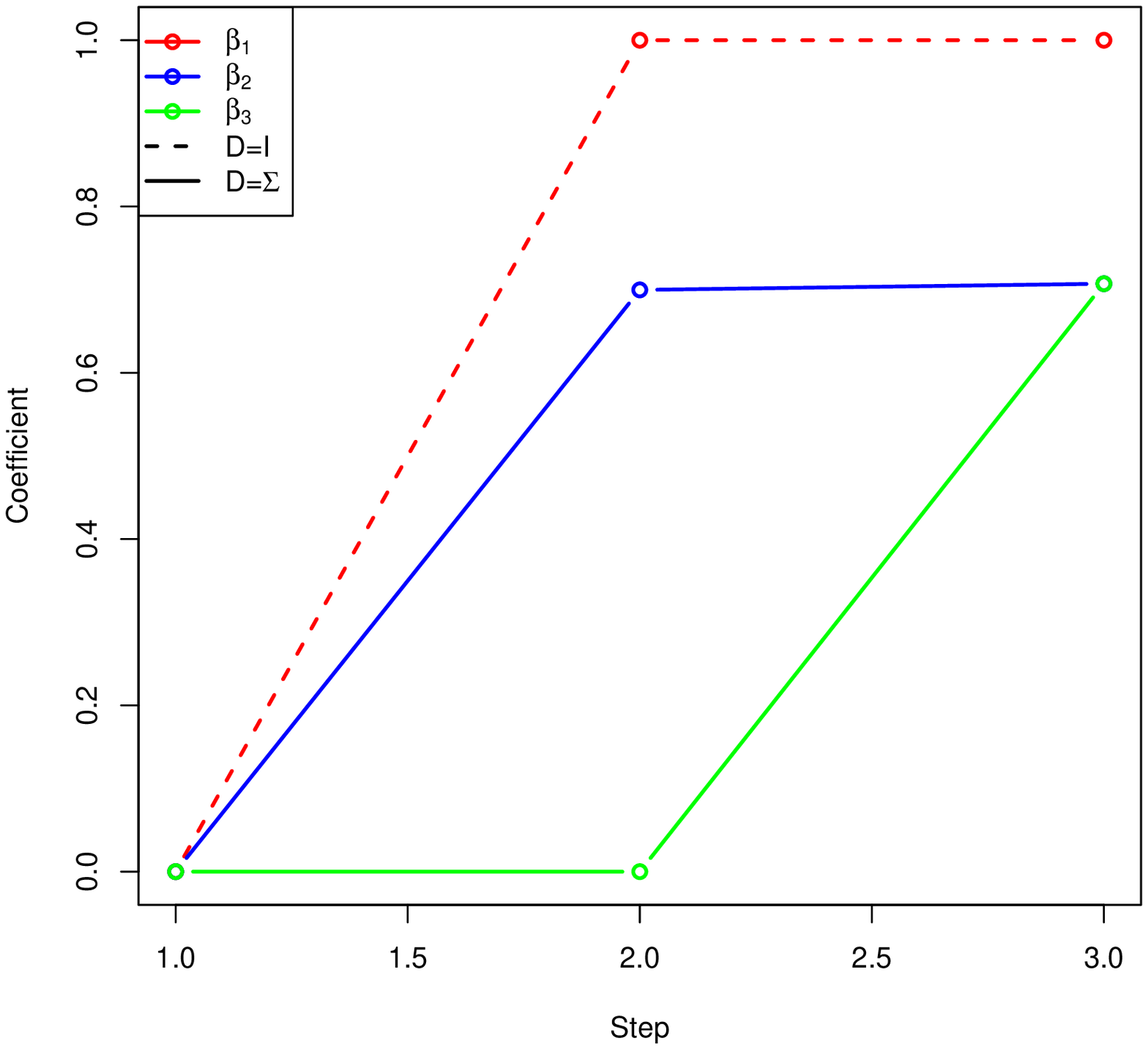}\\
\includegraphics[scale=0.5]{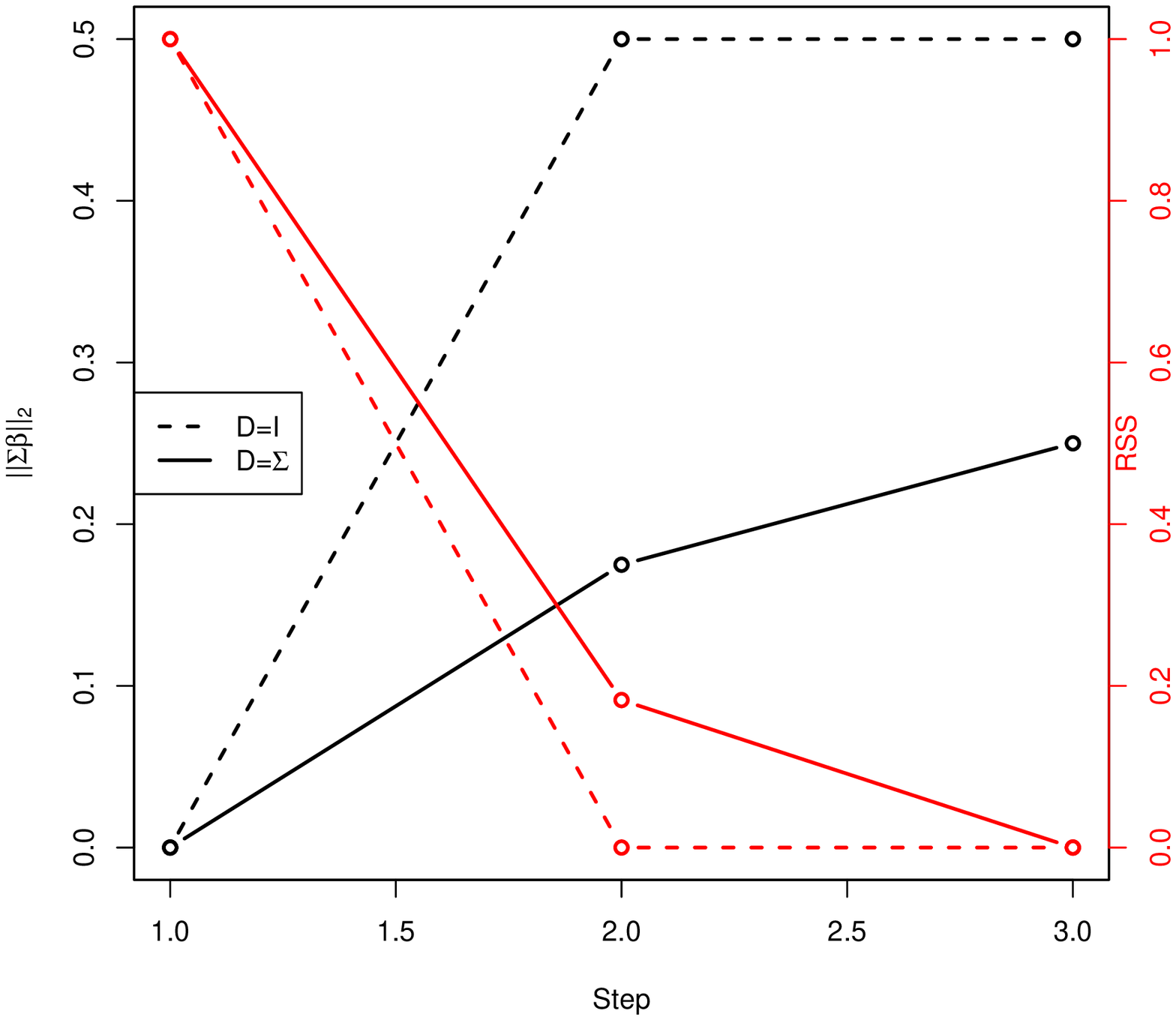} 
\par\end{centering}

\caption{A toy example in which the sparsest solution has more uncertainty
than the next sparsest one (see Section \ref{sub:The-challenge-of-heteroscedastic}).
Left: the regression coefficients at each stage of the FS algorithm.
Only non-zero coefficients are plotted. Right: the uncertainty as
estimated by $||S\boldsymbol{\beta}||_{2}$ (black), with residual
sum of squares (RSS) on the right axis (red). The results are identical
for lasso and LARS.\label{fig:example1}}
\end{figure}

Now consider the change of variables $X\leftarrow XD^{-1}$, $\boldsymbol{\beta}\leftarrow D\boldsymbol{\beta}$,
and $\boldsymbol{\alpha}\leftarrow D\boldsymbol{\alpha}$ as in Section
\ref{sub:Algebraic-interpretation}. In the Dantzig context, this
leads to the linear program: 
\begin{eqnarray*}
\arg\min_{\boldsymbol{\alpha},\boldsymbol{\beta}}\boldsymbol{1}'(D\boldsymbol{\alpha}) & \,\,\mbox{subject to}\,\, & -\boldsymbol{\alpha}\leq\boldsymbol{\beta}\leq\boldsymbol{\alpha}\\
 & \mbox{and} & -\sigma\lambda D\mathbf{1}\leq X'(\mathbf{y}-X\boldsymbol{\beta})\leq\sigma\lambda D\mathbf{1}.
\end{eqnarray*}
Notice that the feasible region is stretched along the noisier dimensions
(proportionally to $\sigma_{\delta_{j}})$, resulting in relaxed requirements
for the residual correlation in those coordinates. This is reasonable,
as we would expect the accuracy for a given variable to be inversely
related to its uncertainty. As in the lasso context, scaling also
results in an explicit $l_{1}$ penalization of the variables commensurate
with their noise level via minimization of the quantity $\boldsymbol{1}'(D\boldsymbol{\alpha})$.
\begin{example*} Continuing the example from Section \ref{sub:The-challenge-of-heteroscedastic},
recall that 
\begin{eqnarray*}
\Sigma_{11}^{2} & = & 1/2,\\
\Sigma_{22}^{2} & = & S_{33}^{2}=1/4,
\end{eqnarray*}
and that the uncertainty in the sparsest solution was greater than
the next sparsest. Figure \ref{fig:example1} gives a concrete illustration
of the solution path for the scaled and unscaled data as well as the
uncertainty in the fit (left panel). After three FS steps (identically
for lasso and LARS), there is zero residual error for both the scaled
and unscaled design (red lines, right panel). However, the uncertainty
associated with $\boldsymbol{\beta}_{2}$ is less than that of $\boldsymbol{\beta}_{1}$
by a factor of 2 (black lines, right panel). \end{example*}

\section{Application to biomass characterization data\label{sec:Application-to-biomass}}

In this section, we present results for both LARS and DS applied to
a biomass characterization data set, with and without scaling. We
highlight the challenges in working with this data, and illustrate
the efficacy of scaling.

\subsection{Description of the data\label{sub:Description-of-the-data}}

\begin{figure}
\begin{centering}
\includegraphics[scale=0.45]{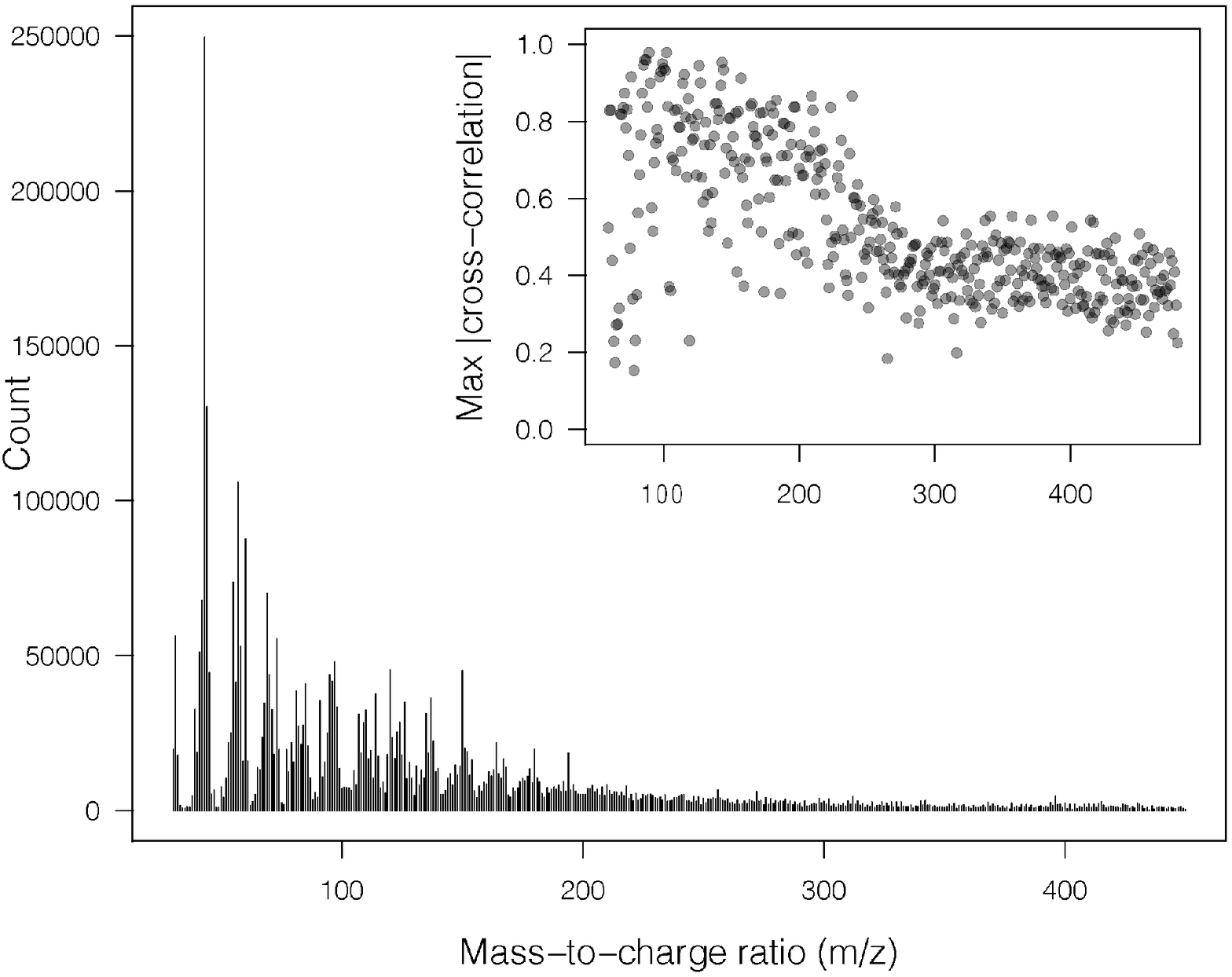}\\
\includegraphics[scale=0.35]{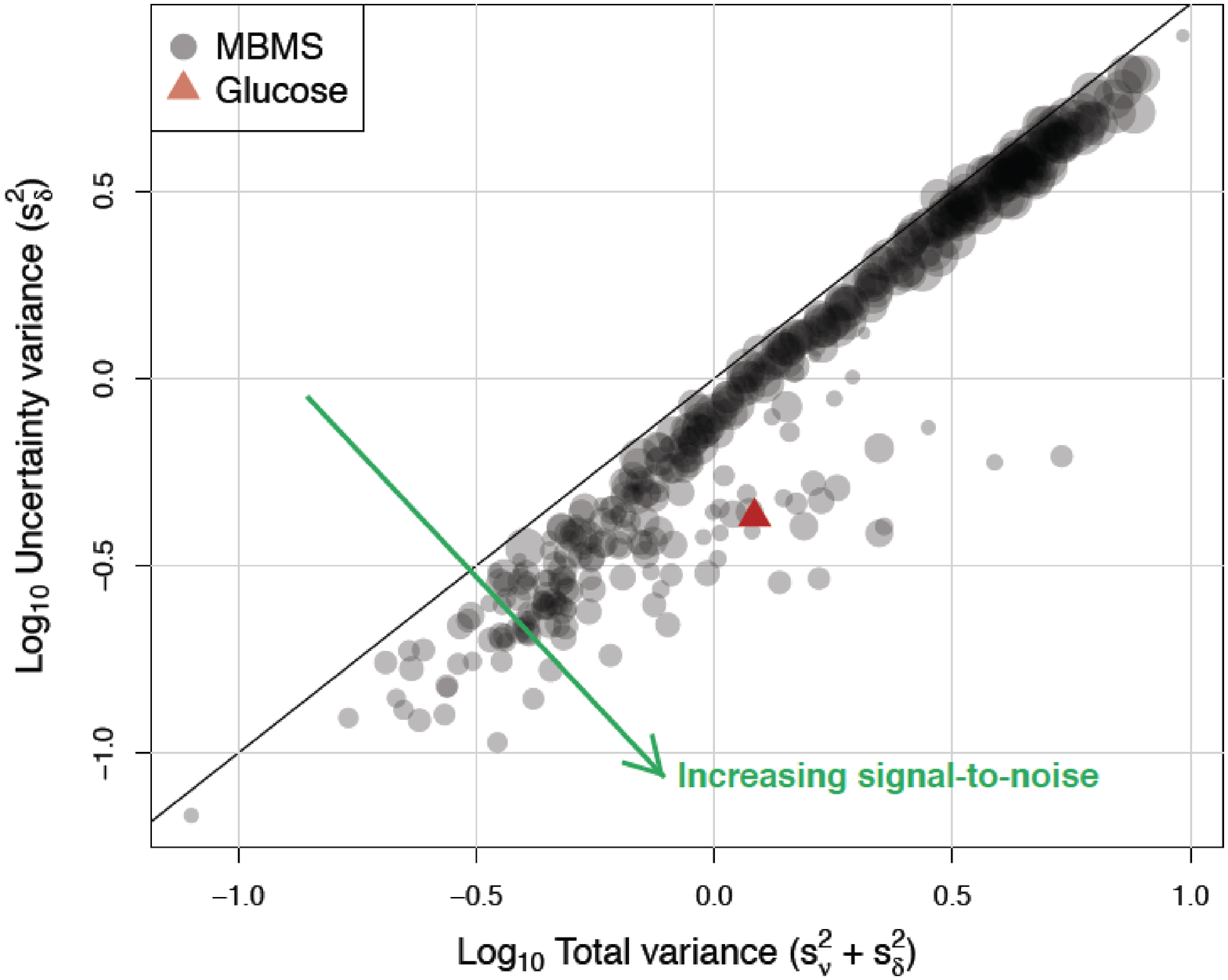} 
\par\end{centering}

\caption{Graphical summary of the biomass characterization data. The top plot
shows a typical raw MBMS spectrum. The inset shows, after pre-processing,
the maximum cross-correlations for each peak and indicates a high
degree of mutual dependence between the predictors. The bottom plot
shows the distribution of variances estimated via a random one-way
ANOVA model before normalization. The marker radius is proportional
to mass-to-charge ratio of the peak, while the response ratio is indicated
by a red triangle. A $1:1$ ratio is indicated by a black line. Points
lying further down and to the right of the solid black line have higher
fidelity (i.e., higher signal-to-noise ratio).\label{fig:mbms-data}}
\end{figure}

The characterization experiment we consider is motivated by a desire
to quickly and inexpensively screen potential biofuel candidates for
recalcitrance - a plant's natural resistance to releasing usable sugars
- after a chemical or enzymatic pre-treatment. Here, $n=759$ switchgrass
plants were grown at different outdoor locations and in uncontrolled
conditions. The predictors consist of $p=421$ pyrolysis molecular
beam mass spectral (pyMBMS or MBMS, \citet{S-Y-N-K-P-D-2009}) lines
measured twice for each sample. As each sample is pyrolyzed, the spectrometer
counts the number of molecules that reach a detector over a range
of mass-to-charge ratios. The raw spectrum for each sample is then
normalized to have unit mass and each peak is divided by a standard
(control) value measured during the same run, allowing samples from
different experiments to be compared directly. So, after pre-processing,
each peak may be thought of as being an expression level for that
mass-to-charge ratio relative to a control.

The response is the mass fraction of extractable glucose as inferred
by the absorbance of 510-nm visible light, where each sample is measured
in triplicate \citep{S-T-L-D-H-D-2011}. In this experiment, a previously
validated linear model is calibrated via measurement of a pure glucose
sample. The mass and absorbance of each biological sample from the
same run are then input to the calibrated model, yielding an estimate
of glucose mass fraction for that sample.

The question we ask is: can the MBMS spectrum (a proxy for chemical
composition) be used to predict the mass fraction of extractable glucose
(usable biofuel)? To answer this, we seek a sparse linear model that
incorporates estimates of uncertainty. Brief justifications for this
approach are: 
\begin{itemize}
\item Sparse: The spectroscopy experiment results in high cross-correlations
between the peaks because large masses break into smaller ones in
a somewhat predictable way. Hence, we expect a significant amount
of redundancy in the peaks. In addition, the relationship between
mass spectral peaks and cell chemistry is complex, making a sparse
model appealing in that it narrows the focus of future investigations
to a few, rather than hundreds, of peaks and their associated compounds. 
\item Incorporates uncertainty: Some of the peaks are far noisier than others,
leading to unequal uncertainties. We would like to ensure that the
model depends on the noisy peaks as little as possible, without completely
excluding them from consideration. 
\item Linear: The assumed physical model is one of linear mixture, \emph{i.e.,}
doubling the concentration of an analyte in the sample should result
in a doubling of its spectral signature. 
\end{itemize}
The data are summarized graphically in Figure \ref{fig:mbms-data}.
A typical raw mass spectrum is shown in the left panel where line
height indicates count following convention for this field. The inset
plot shows the maximum absolute cross-correlation of each peak with
every other peak, from which we infer that there is a high degree
of linear dependence among the variables, especially the smaller masses.
In the right panel, the estimated total and within-sample ANOVA variances
are shown before normalization or scaling, with equality indicated
by a black line. The mass-to-charge ratios of the MBMS lines are proportional
to the marker radius while glucose is indicated by a triangle. Clearly,
many of the peaks are quite noisy, with almost all of the variance
attributed to noise.

\subsection{Methods}

Model selection was performed using nested $k$-fold cross validation
(CV), in which standard $k$-fold CV errors were averaged over $100/k$
outer loops for $k=2,5,$ and $10$. This approach ensured that $100$
different prediction models were validated for each choice of $k$.
We fit both LARS and Dantzig models for comparison. LARS models were
fit in ${\tt R}$ \citet{R-core} using the ${\tt lars}$ package \citep{R-lars}, and Dantzig in ${\tt MATLAB}$ using the ${\tt L_{1}\,\, Homotopy\,\, Toolbox}$
of \citet{MATLAB-2011,L1-HOMOTOPY-MATLAB,ASI-ROM-2010}. $ $ As has
been suggested before (\emph{e.g.}, \citet{ELAD-2010} Chapter 8.5),
it can sometimes be beneficial to regress $\mathbf{y}$ on the sparse
predictor set using another fitting procedure. For comparison, we
used the LARS- and Dantzig-selected peaks as input to cross validated
ridge regression via the ${\tt parcor}$ package in ${\tt R}$ \citep{PARCOR-2010}.
In all instances, the scaling matrix was estimated as part of the
cross validation procedure (see Appendix for details).

\subsection{Results}

\begin{figure}
\begin{centering}
\includegraphics[scale=0.45]{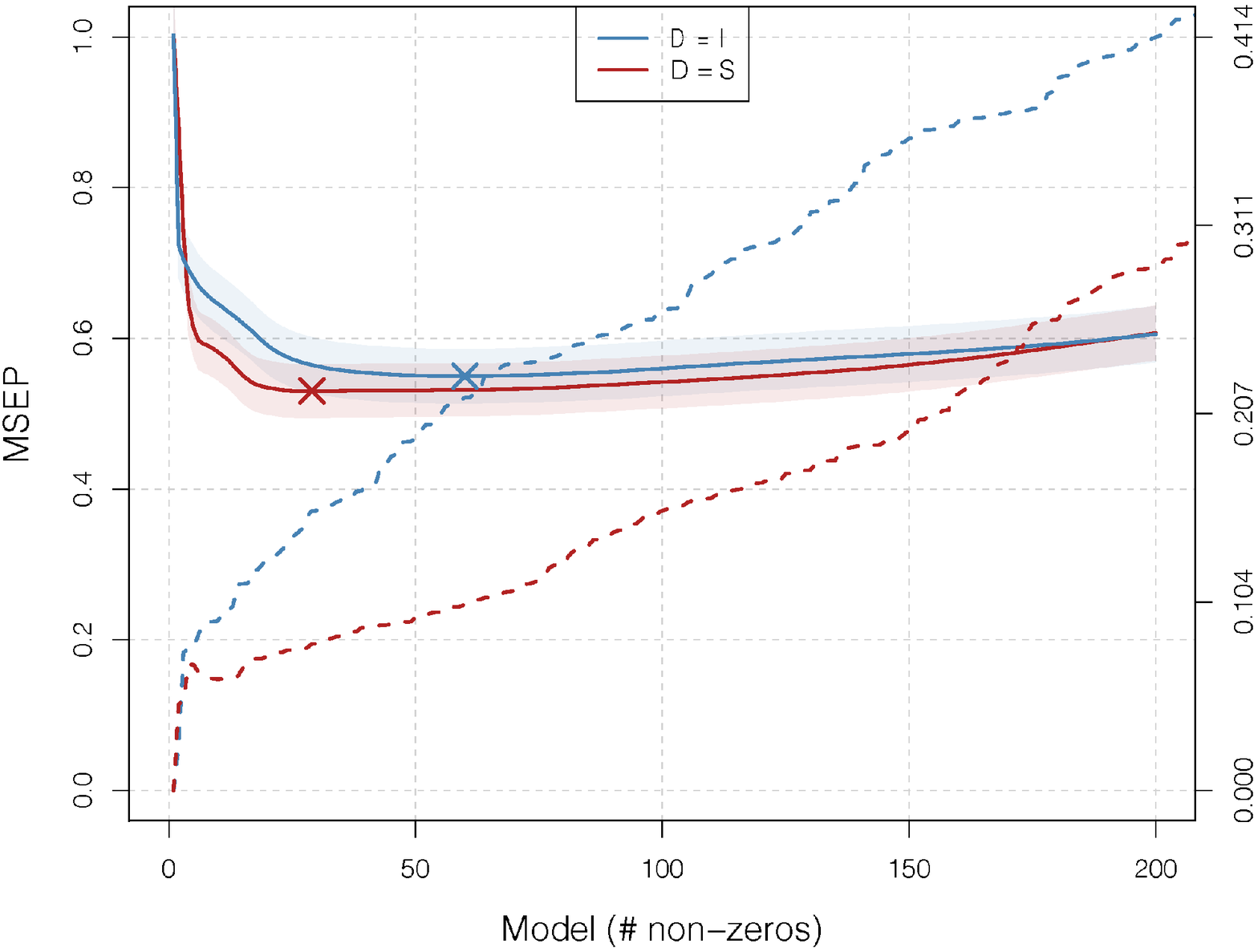}\\
\includegraphics[scale=0.35]{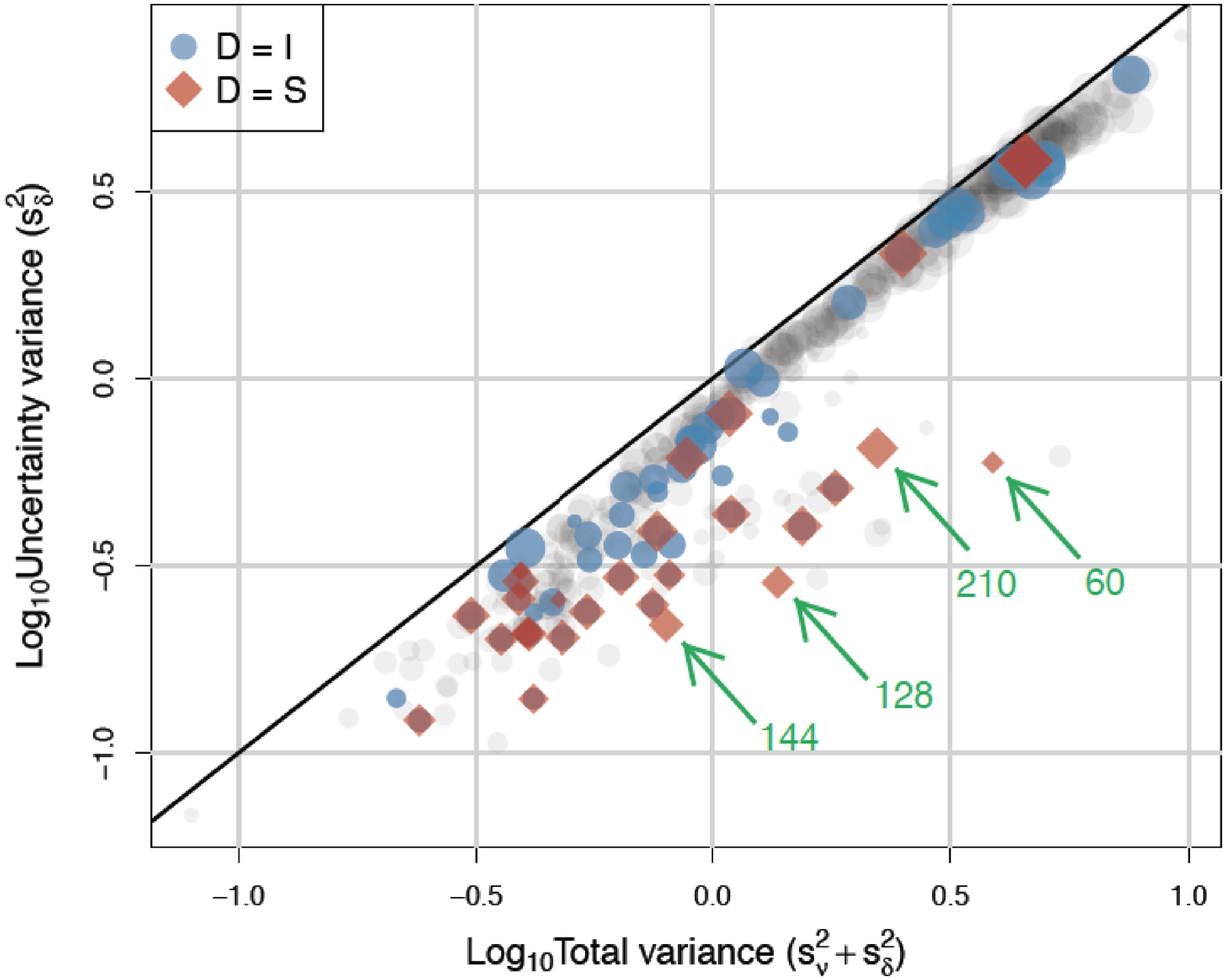} 
\par\end{centering}

\caption{LARS cross validation results for $k=10$. The top plot shows the
mean square error of prediction (MSEP) in bold lines and standard
error as shaded regions. The dashed lines are estimates of uncertainty
via $||S\hat{\boldsymbol{\beta}}||_{2}$, with units on the right
axis. The optimal model is indicated with an x. Apparently, scaling
by uncertainty variances leads to a sparser and more accurate model,
with less associated uncertainty. The bottom plot is identical to the
one in Figure \ref{fig:mbms-data}, but with solution coordinates
selected by LARS given different markers based on scaling (blue circles:
unscaled, red diamonds: scaled). At least 4 peaks with high signal-to-noise
are clearly selected after scaling that are not otherwise (green arrows,
with numbers indicating the $m/z$ ratio). \label{fig:cv-results}}
\end{figure}

\begin{figure}
\begin{centering}
\includegraphics[scale=0.5]{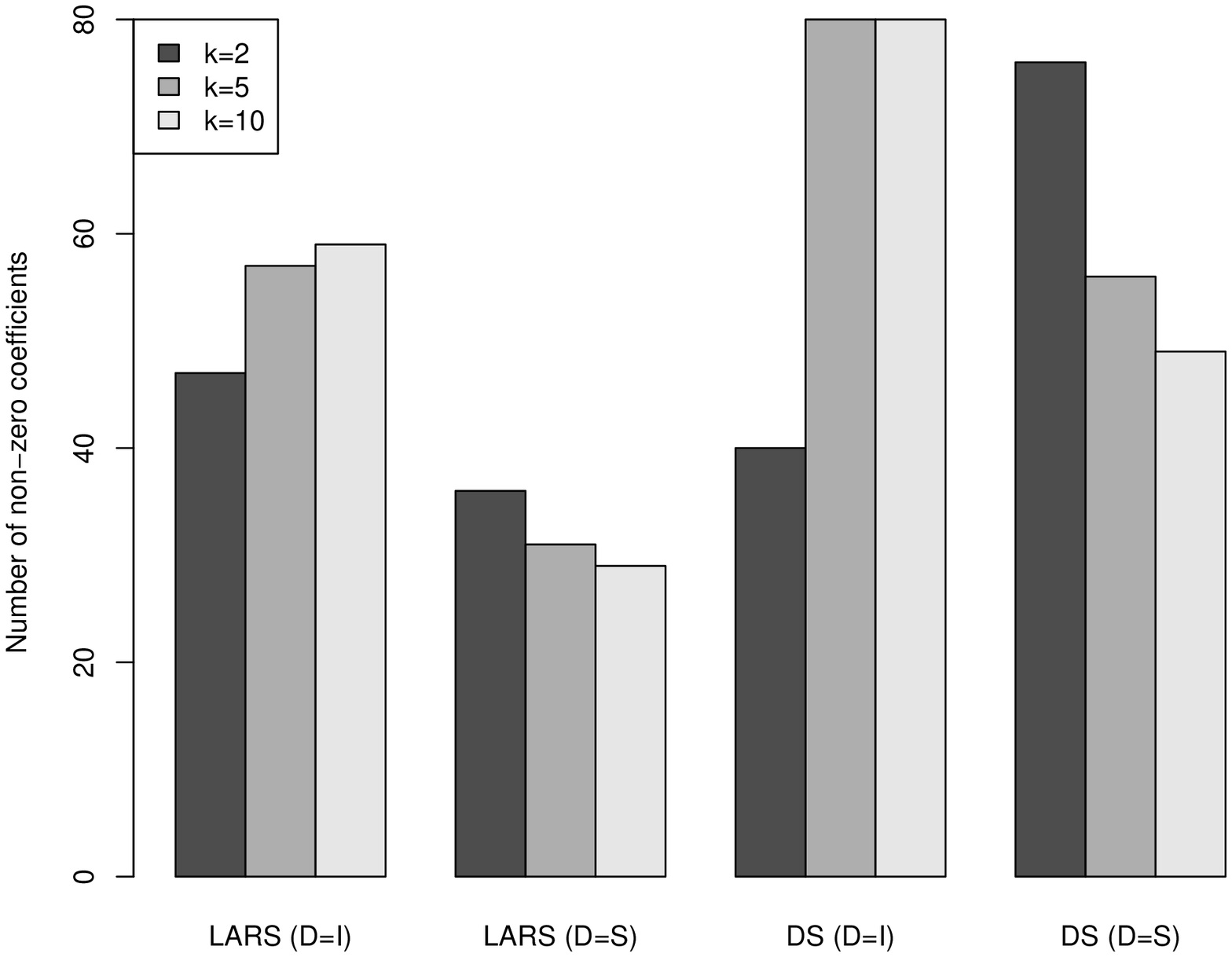} 
\par\end{centering}

\caption{Number of non-zero coefficients for LARS and DS for $k=2,5,10$. In
all cases, the prediction error decreases as $k$ (\emph{i.e.}, the
number of training samples) increases. Without scaling, the more accurate
models use more variables while with scaling, remarkably, they use
fewer. \label{fig:cv-sparsity}}
\end{figure}

Cross-validation results are given in Table \ref{tab:k-fold-cross-validation}
in the Appendix, and may be summarized as follows. Scaling leads to: 
\begin{enumerate}
\item improved accuracy, as measured by cross-validated MSEP, for both LARS
and DS (Figure \ref{fig:cv-results}) 
\item increased sparsity for both LARS and DS (Figure \ref{fig:cv-sparsity}) 
\item higher degree of consistency between LARS and DS 
\end{enumerate}

The left panel of Figure \ref{fig:cv-results} shows the prediction
error per LARS step (solid lines), the standard error (shaded regions),
and the uncertainty as estimated by $||S\hat{\boldsymbol{\beta}}||_{2}$
for $k=10$ (dashed lines). The optimal models are indicated by x's.
While the standard error of prediction is similar for the scaled and
unscaled case, the uncertainty accumulates more slowly for the scaled
input (almost identical results hold for $k=2,5$, not shown). The
right panel provides a graphical impression of the quality of the
variables selected for the scaled and unscaled data. One can see that,
in general, the scaled approach (red diamonds) leads to selection
of peaks with higher signal-to-noise ratio, indicated by green arrows,
than the unscaled (blue circles).

Figure \ref{fig:cv-sparsity} shows the sparsity of the LARS and DS
models as a function of the CV fold sizes. Remarkably, the number
of non-zero coefficients for both LARS and DS actually increases with
increasing sample size when the data are unscaled. This is somewhat
surprising since, heuristically, one would expect the model selection
to be more discriminating as more samples are utilized. On the other
hand, the number of non-zeros decreases with increasing sample size
for the scaled data. To explain this, we speculate that when the data
are unscaled, it is more likely for the algorithm to select variables
that are either neutral or even detrimental with respect to prediction.
If this is the case, then our results suggest that scaling leads to
a more discriminating variable selection and higher prediction accuracy.

Finally, while the LARS and DS solutions are not in perfect agreement
in either case (scaled or unscaled), they are seen to be in better
agreement after scaling. Only $46\%$ of the LARS peaks are also selected
by DS without scaling, while the number is $86\%$ with scaling. Alternatively,
of the total number of distinct peaks selected by LARS and DS combined,
only 24\% are common to both without scaling, while 45\% are common
to both with scaling.

\subsection{Discussion}

It should be stated up front that the assumption of linearity made
in Section \ref{sub:Description-of-the-data} does not appear to be
completely valid. While the assumption should be valid on physical
grounds, there are obviously experimental, biological or other factors
that introduce significant error terms beyond those formulated in
Section \ref{sub:Model}. That said, the fact that scaling leads to
a reduction in CV error, increased sparsity, and better agreement
between LARS and DS suggests that the method can still be practically
useful under non-ideal circumstances.

While some of the peaks identified by both scaled LARS and DS have
been previously recognized as related to recalcitrance, many have
not (see Table \ref{tab:peaks-identified} in Appendix). Of particular
interest are the peaks with large $m/z$ ratios, as these are less
likely to be correlated coincidentally with recalcitrance: light particles
can originate from a variety of sources, but less so for larger particles.
Furthermore, some of the unknown peaks have regression coefficients
that are not small. We believe that these results warrant taking a
further look at the unknown peak associations to better understand
chemical mechanisms of recalcitrance.

\section{Conclusions}

We have argued that sparse regression under design uncertainty presents
several challenges that (to the best of our knowledge) have not been
addressed in the literature. Focusing on the the uncertainty term,
$||\Sigma\hat{\boldsymbol{\beta}}||_{2}^{2}$, in the residual error
(\ref{eq:residual-error}), we propose a scaling of the design variables
by their uncertainty variances. In the context of greedy algorithms,
doing this guarantees a uniform growth of uncertainty regardless of
the order in which the variables are selected. Within the lasso formulation,
scaling is shown to enforce an $l_{1}$ penalization of the uncertainty.
In the Dantzig selector context, scaling leads to modified bounds
on the residual error that reflect the amount of uncertainty associated
with each variable.

In a biomass characterization application, scaling is shown empirically
to reduce uncertainty in the optimal solution. It also leads to sparser
solutions and lower prediction error. The solution estimates are improved
even further if the LARS- and Dantzig-selected peaks are used independently
for ridge regression. In addition, these models are more consistent
with one another after scaling, that is, they identify more of the
same predictors. The improvements resulting from scaling are promising
and deserve further consideration.

\section*{Acknowledgements}

The authors would like to thank Terry Haut for many useful conversations,
Peter Graf for his critical eye, and Matthew Reynolds for help with
proofreading. This work was supported by the DOE Offi{}ce of Biological
and Environmental Research through the BioEnergy Science Center (BESC).
BioEnergy Science Center (BESC) is a US Department of Energy Bioenergy
Research Center supported by the Offi{}ce of Biological and Environmental
Research in the DOE Offi{}ce of Science.

\newpage{}

\section*{Appendix\label{sec:Appendix}}

\subsubsection*{ANOVA model.}

We use a one way, random effects ANOVA model to estimate the uncertainty
variances. Let $r$ denote the number of replicate measurements of
a random variable $Z$ and let $n$ denote the number of samples.
The relevant quantities needed to estimate the variance components
are shown in Table \ref{tab:ANOVA}. In particular, the estimates
are $s_{\delta}^{2}$ = SSE/df, and $s_{\nu}^{2}$ = (SSTr/df-$\sigma_{\delta}^{2}$)/$r$.

\begin{table}[h]
\begin{centering}
\begin{tabular}{|c|c|c|c|}
\hline 
Source  & df  & Sum of squares  & Expected mean square\tabularnewline
\hline 
\hline 
Treatment  & $n-1$  & SSTr = $\sum_{i=1}^{n}r(\bar{z}_{i\cdot}-\bar{z}_{\cdot\cdot})^{2}$  & $r\sigma_{\nu}^{2}+\sigma_{\delta}^{2}$\tabularnewline
\hline 
Error (uncertainty)  & $n(r-1)$  & SSE = $\sum_{i=1}^{n}\sum_{j=1}^{r}(z_{ij}-\bar{z}_{i\cdot})^{2}$  & $\sigma_{\delta}^{2}$\tabularnewline
\hline 
\end{tabular}
\par\end{centering}

\caption{Standard one-way, random effects ANOVA table.\label{tab:ANOVA}}
\end{table}

\subsubsection*{Cross validation procedure.}

For clarity, we outline our procedure for cross-validated model selection
using replicated measurements. It is a completely standard cross-validation
procedure with the simple addition that we estimate the uncertainty
variances only from the training data.

For each of the $k$ cross-validation groups: 
\begin{enumerate}
\item Split the data into training, $\{\mathbf{y}_{train},X_{train}\}$,
and test sets, $\{\mathbf{y}_{test},X_{test}\}$, of appropriate sizes. 
\item Using only $X_{train}$, estimate the error variances, $s_{\delta_{j}}^{2}$,
via a suitable method (we used one-way, random effects ANOVA). 
\item Form the diagonal matrix $D_{jj}=s_{\delta_{j}}^{2}$ and scale
the training data, $X_{train}\leftarrow X_{train}D^{-1}$. 
\item Fit the desired models to $\mathbf{y}_{train}$ using scaled $X_{train}$. 
\item Using $D$ from step 3, scale the test data, $X_{test}\leftarrow X_{test}D^{-1}$,
and predict. 
\[
\]

\end{enumerate}
\begin{sidewaystable}
\begin{centering}
\begin{tabular}{|l|l|c|c|c|c|c|}
\hline 
Model selection method  & Scaling  & $k$  & \# predictors  & MSEP  & Avg \# predictors  & Avg MSEP\tabularnewline
\hline 
\hline 
 &  & 2  & 47  & 0.564  &  & \tabularnewline
\hline 
LARS  & NO  & 5  & 57  & 0.551  & 54.3  & 0.555\tabularnewline
\hline 
 &  & 10  & 59  & 0.550  &  & \tabularnewline
\hline 
 &  & 2  & 36  & 0.539  &  & \tabularnewline
\hline 
LARS  & YES  & 5  & 31  & 0.531  & 31.7  & 0.533\tabularnewline
\hline 
 &  & 10  & 28  & 0.530  &  & \tabularnewline
\hline 
 &  & 2  & 47  & 0.508  &  & \tabularnewline
\hline 
LARS-RR  & NO  & 5  & 57  & 0.485  & 54.3  & 0.493\tabularnewline
\hline 
 &  & 10  & 59  & 0.485  &  & \tabularnewline
\hline 
 &  & 2  & 36  & 0.492  &  & \tabularnewline
\hline 
LARS-RR  & YES  & 5  & 31  & 0.491  & 31.7  & 0.492\tabularnewline
\hline 
 &  & 10  & 28  & 0.494  &  & \tabularnewline
\hline 
 &  & 2  & 40  & 0.622  &  & \tabularnewline
\hline 
Dantzig selector  & NO  & 5  & 80  & 0.583  & 66.7  & 0.599\tabularnewline
\hline 
 &  & 10  & 80  & 0.574  &  & \tabularnewline
\hline 
 &  & 2  & 76  & 0.547  &  & \tabularnewline
\hline 
Dantzig selector  & YES  & 5  & 56  & 0.528  & 60.3  & 0.536\tabularnewline
\hline 
 &  & 10  & 49  & 0.523  &  & \tabularnewline
\hline 
 &  & 2  & 40  & 0.512  &  & \tabularnewline
\hline 
Dantzig-RR  & NO  & 5  & 80  & 0.487  & 66.7  & 0.494\tabularnewline
\hline 
 &  & 10  & 80  & 0.482  &  & \tabularnewline
\hline 
 &  & 2  & 76  & 0.460  &  & \tabularnewline
\hline 
Dantzig-RR  & YES  & 5  & 56  & 0.454  & 60.3  & 0.455\tabularnewline
\hline 
 &  & 10  & 49  & 0.451  &  & \tabularnewline
\hline 
 &  & 2  & 421  & 0.533  &  & \tabularnewline
\hline 
Ridge regression  & NO  & 5  & 421  & 0.536  & 421  & 0.535\tabularnewline
\hline 
 &  & 10  & 421  & 0.535  &  & \tabularnewline
\hline 
 &  & 2  & 421  & 0.515  &  & \tabularnewline
\hline 
Ridge regression  & YES  & 5  & 421  & 0.517  & 421  & 0.516\tabularnewline
\hline 
 &  & 10  & 421  & 0.517  &  & \tabularnewline
\hline 
\end{tabular}
\par\end{centering}

\caption{Results of $k$-fold cross validation (see also Figure \ref{fig:cv-sparsity}).
The -RR suffix indicates ridge regression was performed on the subset
selected by the corresponding sparse algorithm. For comparison, results
for ridge regression using all of the predictor variables are also
shown.\label{tab:k-fold-cross-validation}}
\end{sidewaystable}

\begin{table}
\begin{centering}
\begin{tabular}{|c|c|c|}
\hline 
$m/z$  & Assignment in\citet{S-K-T-F-D-2008}  & Avg. coefficient (relative to max)\tabularnewline
\hline 
\hline 
45  & ?  & \texttt{+1.0000}\tabularnewline
\hline 
60  & C5 sugars  & \texttt{-0.0534}\tabularnewline
\hline 
120  & Vinylphenol  & \texttt{-0.2519}\tabularnewline
\hline 
126  & ?  & \texttt{-0.1432}\tabularnewline
\hline 
128  & ?  & \texttt{-0.0482}\tabularnewline
\hline 
129  & ?  & \texttt{-0.2177}\tabularnewline
\hline 
137  & Ethylguaiacol, homovanillin, coniferyl alcohol  & \texttt{-0.7648}\tabularnewline
\hline 
143  & ?  & \texttt{+0.1795}\tabularnewline
\hline 
144  & ?  & \texttt{-0.2338}\tabularnewline
\hline 
150  & Vinylguaiacol  & \texttt{+0.8777}\tabularnewline
\hline 
159  & ?  & \texttt{+0.1016}\tabularnewline
\hline 
160  & ?  & \texttt{+0.1988}\tabularnewline
\hline 
164  & Allyl $\pm$ propenyl guaiacol  & \texttt{-0.3966}\tabularnewline
\hline 
168  & 4-Methyl-2, 6-dimethoxyphenol  & \texttt{-0.3949}\tabularnewline
\hline 
175  & ?  & \texttt{+0.0554}\tabularnewline
\hline 
182  & Syringaldehyde  & \texttt{-0.0734}\tabularnewline
\hline 
194  & 4-Propenylsyringol  & \texttt{-0.8661}\tabularnewline
\hline 
208  & Sinapyl aldehyde  & \texttt{-0.0038}\tabularnewline
\hline 
210  & Sinapyl alcohol  & \texttt{+0.4334}\tabularnewline
\hline 
226  & ?  & \texttt{+0.3021}\tabularnewline
\hline 
264  & ?  & \texttt{+0.1883}\tabularnewline
\hline 
287  & ?  & \texttt{-0.1646}\tabularnewline
\hline 
371  & ?  & \texttt{-0.1970}\tabularnewline
\hline 
374  & ?  & \texttt{+0.2153}\tabularnewline
\hline 
\end{tabular}
\par\end{centering}

\caption{Peaks identified by both scaled LARS and DS for $k=10$. The peaks
previously identified in \citet{S-K-T-F-D-2008} as significant to
sugar release are described where possible. The LARS and DS regression
coefficients were averaged and divided by the maximum to highlight
the relative significance and sign of correlation of the peaks. Some
of the most highly-weighted variables have not previously been identified
as being related to recalcitrance.\label{tab:peaks-identified}}
\end{table}

\newpage{}

\bibliographystyle{plainnat}
\bibliography{refs-aoas}

\end{document}